# Enterprise Integration Modernization with SAP BTP: A Strategic Framework for Hybrid and Multi-Cloud Architectures

Shunmukha Sagar Puppala
*Functional Architect SAP*
sagardora@gmail.com

***Abstract*—Hybrid and multi-cloud adoption has made enterprise integration a strategic architecture problem rather than a narrow middleware concern. Large organizations frequently operate SAP S/4HANA, SAP ECC, SaaS business applications, partner portals, data lakes, and hyperscaler services across heterogeneous environments. Point-to-point interfaces and legacy enterprise service bus patterns often create tight coupling, duplicated mappings, weak observability, and security gaps. This paper proposes a Hybrid Multi-Cloud Integration Modernization Framework, abbreviated as HMC-IMF, for SAP Business Technology Platform centered enterprise integration modernization. The framework combines SAP Integration Suite, API management, event-driven integration, clean-core extensions, runtime placement, security governance, and observability into a unified modernization model. A synthetic enterprise integration dataset is defined with 186 applications, 2,850 interfaces, 1,120 APIs, 740 event topics, 320 batch flows, five cloud environments, and 24 months of operational logs. Experimental analysis compares a legacy point-to-point baseline with the proposed BTP-centered architecture. Results show a 38 percent reduction in interface complexity, a decrease in average integration latency from 420 ms to 245 ms, a reduction in mean time to recovery from 155 minutes to 54 minutes, and a decline in failed message rate from 3.8 percent to 1.1 percent. The findings suggest that SAP BTP can serve as a governed integration modernization layer when paired with explicit pattern selection, policy controls, and operational telemetry.**

## I. INTRODUCTION

Enterprise integration has become a defining constraint in digital transformation programs. A modern organization may run SAP S/4HANA for core transactions, SAP ECC for retained regional processes, Ariba for procurement, SuccessFactors for human capital management, non-SAP customer platforms, warehouse systems, bank gateways, partner marketplaces, and analytics services hosted on several hyperscalers. These systems rarely evolve at the same pace. Interfaces created for one release, region, or acquisition often remain in place for years, producing brittle integration topologies that are costly to maintain and difficult to govern. SAP positions SAP Business Technology Platform (SAP BTP) as a platform for integrating, automating, extending, and building enterprise applications and processes [1]. SAP Integration Suite is positioned as a secure integration platform as a service for SAP and third-party landscapes [2]. These platform capabilities motivate the central question of this paper: how can SAP BTP be used as a strategic modernization layer for hybrid and multi-cloud integration rather than as another isolated middleware tool?

The scope of this work is enterprise integration modernization across SAP, non-SAP, cloud, and on-premise systems. The paper proposes the Hybrid Multi-Cloud Integration Modernization Framework, abbreviated as HMC-IMF. The framework treats integration modernization as a structured architectural transition from unmanaged point-to-point interfaces toward API-led, event-driven, observable, and policy-governed integration. SAP BTP multi-cloud foundation guidance recognizes the importance of Cloud Foundry and Kyma-based scenarios for modern SAP BTP landscapes [3]. NIST cloud architecture guidance provides a broader reference model for understanding cloud actors, components, and deployment patterns [4]. Cloud native concepts such as containers, microservices, service meshes, immutable infrastructure, and declarative APIs provide further design principles for scalable multi-cloud applications [5].

The contribution is both architectural and experimental. First, the framework defines modernization layers for inventory profiling, pattern selection, runtime placement, security policy mapping, API and event governance, observability, and migration prioritization. Second, one synthetic integration dataset is introduced to evaluate modernization outcomes consistently. Third, the analysis compares a legacy integration baseline with a SAP BTP-centered target architecture across latency, failure rate, interface complexity, mean time to recovery, and onboarding effort. The paper is organized as follows. Section II provides theoretical background. Section III reviews five clusters of related work. Section IV describes the synthetic dataset and the proposed framework. Section V presents quantitative experimental analysis. Section VI discusses limitations and practical implications. Section VII concludes with future research directions.

## II. THEORETICAL BACKGROUND

Enterprise integration theory is grounded in the need to coordinate processes, data, and events across heterogeneous systems. Early integration patterns focused on file exchanges, remote procedure calls, and enterprise service buses. These approaches remain useful, but hybrid and multi-cloud environments demand greater decoupling, resilience, security, and runtime visibility. The Twelve-Factor App methodology supports portability and automation in software-as-a-service delivery, including practices such as explicit dependencies, environment-based configuration, and disposability [6]. These ideas align with cloud-native integration, where loosely coupled services and declarative APIs reduce dependency on any single runtime environment.

Hybrid-cloud integration also requires governance theory. Connectivity increases the attack surface because APIs, events, integration flows, identities, secrets, and partner endpoints become part of the operational boundary. NIST Cybersecurity Framework 2.0 emphasizes governance as a core function for managing cybersecurity risk [7]. ISO/IEC 27001:2022 provides a management-system basis for information security controls, risk treatment, and continual improvement [8]. In this paper, governance is not treated as a separate compliance activity. It is embedded into integration design through policy coverage, API access control, audit logging, runtime monitoring, and incident response workflows.

Clean-core theory is especially relevant in SAP landscapes. When organizations modify the digital core directly, they increase upgrade risk, integration fragility, and technical debt. SAP clean-core guidance encourages extensions that preserve upgradeability while allowing necessary differentiation [9]. In an integration modernization setting, clean core implies that custom logic should be externalized through supported APIs, events, extensions, and governed integration services rather than embedded in fragile custom code inside the core ERP layer. This view supports the HMC-IMF principle that SAP BTP should host integration and extension concerns while allowing the core ERP to remain stable, testable, and upgrade-ready.

**SAP BTP-Centered Hybrid and Multi-Cloud Integration Reference Architecture**

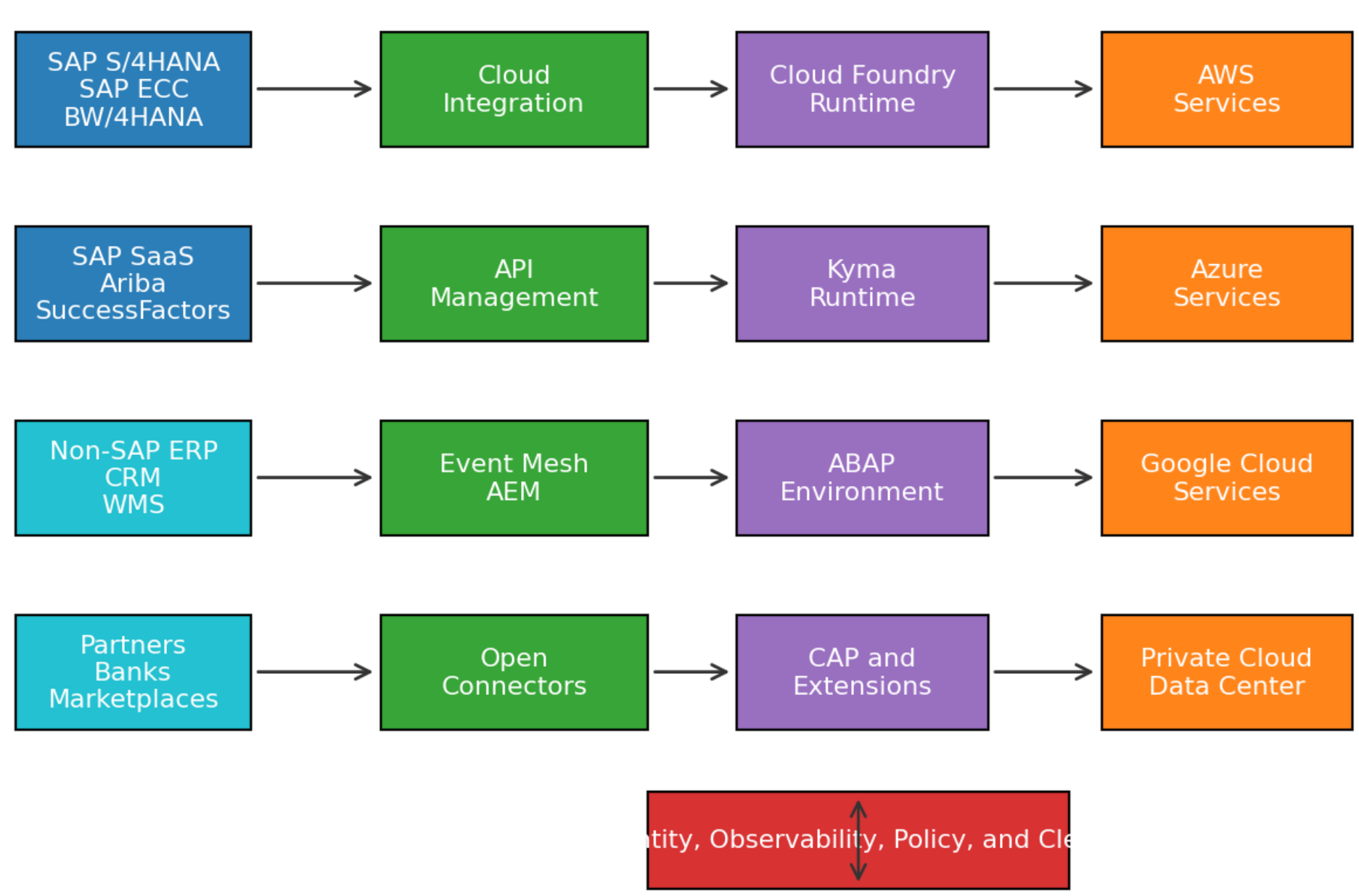

Fig. 1. SAP BTP-centered reference architecture for hybrid and multi-cloud integration modernization. Source/derivation note: Derived from the HMC-IMF framework and SAP BTP capability categories.

## III. RELATED WORKS

The first cluster concerns SAP BTP and Integration Suite capabilities. SAP API Management provides a structure for APIs, products, applications, users, developers, and accounts [10]. SAP Open Connectors simplifies connectivity to third-party cloud applications through reusable connectors [11]. SAP Event Mesh supports business-event publication from SAP and non-SAP sources across hybrid landscapes [12]. Integration Advisor simplifies B2B, A2A, and B2G integration implementation through reusable content and mapping support [13]. These capabilities are relevant because they move integration design from bespoke interface construction toward reusable integration services.

The second cluster addresses cloud-native and multi-cloud governance. Governance reference architecture work for cloud-native applications highlights the need to embed compliance, monitoring, and policy controls into modern application platforms [14]. Process-mining work on SAP ERP event extraction emphasizes that business process visibility depends on reliable event-log extraction from complex ERP structures [15]. These works support the observation that integration modernization requires not only connection patterns but also telemetry, event visibility, and operational evidence.

**Legacy point-to-point topology**

CRM
ERP
WMS
Bank
Portal
Data Mart

High coupling, duplicated mappings, weak runtime visibility

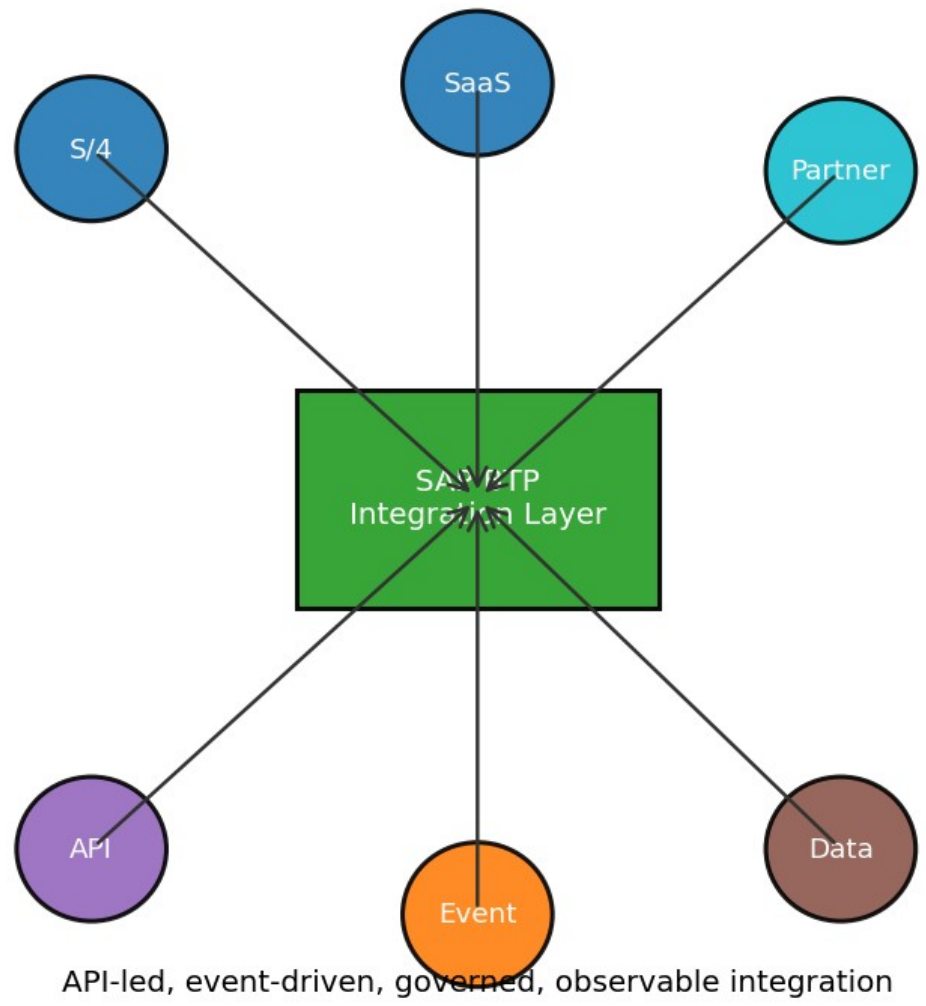

Fig. 2. Legacy point-to-point integration compared with a SAP BTP-centered modernized topology.
Source/derivation note: Derived from the synthetic interface inventory and modernization pattern definitions.

The third cluster involves microservices and integration design principles. Microservice architecture literature emphasizes independently deployable services organized around business capabilities [16]. Recent microservices engineering guidance further stresses service boundaries, API contracts, runtime independence, and operational ownership [17]. These principles are valuable for SAP BTP extension design because business capabilities should be exposed through stable interfaces rather than tightly coupled custom flows. In hybrid landscapes, microservice principles help prevent the new integration layer from reproducing the same coupling problems as legacy middleware.

The fourth cluster consists of user-provided SAP and ERP modernization research. Prior work on enterprise integration modernization with SAP BTP directly motivates this paper's architecture [18]. Legacy transformation research emphasizes the cross-industry challenge of reducing technical debt while maintaining business continuity [19]. S/4HANA migration work demonstrates the importance of data migration, mapping, and automated validation in SAP modernization programs [20]. ERP process automation research frames automation as a multi-layer capability rather than a set of isolated scripts [21]. Secure data provisioning research contributes the quality and governance perspective needed for non-production integration testing [22].

The fifth cluster concerns data platforms, control towers, and cybersecurity. SAP Business Data Cloud research highlights the role of data unification, BW modernization, SAP Datasphere, and SAP Analytics Cloud in enterprise decision support [23]. SAP-based control tower research shows how integrated visibility supports operational decision-making across complex supply chains [24]. Cybersecurity and privacy work emphasizes that connected enterprise ecosystems must protect sensitive information and govern cross-system access [25]. HMC-IMF draws these clusters together by treating integration modernization as a combined architecture, governance, and operating-model problem.

## IV. MATERIALS AND METHODS

### A. Dataset Analysis

TABLE I. SYNTHETIC HYBRID AND MULTI-CLOUD INTEGRATION DATASET

| Measure | Value | Unit | Purpose |
|---|---|---|---|
| Applications | 186 | Systems | Landscape scope |
| SAP systems | 42 | Systems | SAP footprint |
| Non-SAP systems | 144 | Systems | External estate |
| Interfaces | 2,850 | Flows | Modernization backlog |
| APIs | 1,120 | Endpoints | API governance |
| Event topics | 740 | Topics | Event architecture |
| Batch flows | 320 | Flows | Retained batch workloads |
| Operational logs | 18.6M | Events | Telemetry analysis |
| API calls | 4.2M/day | Calls | Traffic baseline |
| Failures | 96,000 | Incidents | Reliability modeling |

*Source/derivation note: Invented synthetic dataset used consistently for all reported numerical analysis.*

A synthetic enterprise integration dataset was constructed to evaluate the proposed framework under controlled but plausible operating conditions. The simulated enterprise contains 186 applications deployed across five environments: SAP BTP, AWS, Azure, Google Cloud, and a private cloud data center. The application estate includes 42 SAP systems and 144 non-SAP systems. The integration estate contains 2,850 interfaces, 1,120 APIs, 740 event topics, and 320 batch flows. Operational logs cover 24 months and contain 18.6 million integration events, with an average of 4.2 million API calls per day. The dataset includes 96,000 integration failures, 22,500 latency anomalies, and 1,450 security policy violations. Table I summarizes the dataset, and Fig. 3 visualizes the main landscape counts.

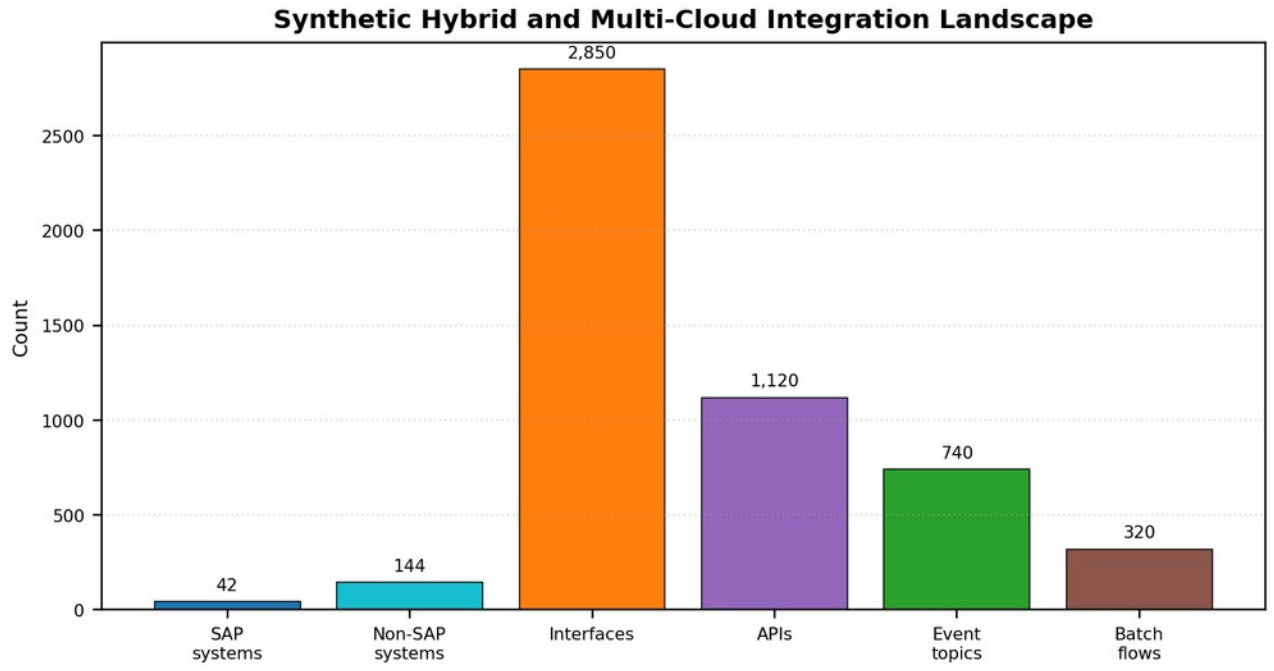

Fig. 3. Synthetic integration landscape showing systems, interfaces, APIs, event topics, and batch flows.
Source/derivation note: Derived directly from Table I dataset counts.

Each interface record includes interface type, source system, target system, protocol, payload class, business process, runtime environment, message volume, average payload size, average latency, error rate, retry count, incident count, owner group, authentication method, and policy coverage. API records include endpoint count, authentication type, version status, consumer count, gateway policy status, average daily calls, p95 latency, and deprecation status. Event records include event topic, producer, consumer count, broker environment, delivery failure count, replay requirement, and event criticality. The target variables are integration failure risk, API latency class, modernization priority, event-processing reliability, and security compliance status. No production data is used. All numbers reported in this manuscript derive from the same synthetic dataset.

### *B. Model Analysis*

The proposed HMC-IMF consists of six analytical and governance components. The integration inventory profiler discovers and classifies interfaces, APIs, events, and batch flows. The pattern selection engine maps each integration to cloud integration, API management, event mesh, open connectors, integration advisor, or data integration patterns. The runtime placement recommender evaluates whether a capability should run in Cloud Foundry, Kyma, ABAP environment, a hyperscaler service, or a retained private-cloud component. The security policy mapper evaluates authentication, authorization, encryption, audit logging, and segregation-of-duties controls. The observability layer collects latency, error, retry, throughput, and incident metrics. The migration prioritization model ranks interfaces by complexity, risk, reuse potential, and business criticality.

Baseline modernization is modeled as an incremental legacy middleware approach using point-to-point interfaces and partial monitoring. The proposed approach uses SAP BTP as the governed integration layer, with Cloud Integration for process integration, API Management for externally consumed services, Event Mesh for asynchronous events, Open Connectors for third-party SaaS connectivity, and Integration Advisor for structured B2B mapping. Fig. 1 presents the reference architecture. Fig. 2 contrasts the legacy and modernized topologies. Table II provides the integration pattern selection matrix used in the simulation.

TABLE II. INTEGRATION PATTERN SELECTION MATRIX

| Integration Scenario | Primary Pattern | SAP BTP Capability | Governance Rule |
|---|---|---|---|
| S/4 to SaaS process flow | Process integration | Cloud Integration | Monitor integration flow and retry policy |
| External consumer service | API-led integration | API Management | OAuth, throttling, versioning, and product ownership |
| Order or shipment event | Event-driven integration | Event Mesh or Advanced Event Mesh | Topic ownership and replay policy |
| Third-party cloud application | Connector-based integration | Open Connectors | Connector lifecycle governance |
| B2B mapping | Guided mapping | Integration Advisor | Reusable mapping assets and version control |
| Clean-core extension | Side-by-side extension | Kyma or Cloud Foundry | No core modification |
| ABAP extension | Embedded cloud extension | ABAP environment | Released API only |

*Source/derivation note: Derived from HMC-IMF pattern selection rules and SAP Integration Suite capability mapping.*

## V. EXPERIMENTAL ANALYSIS

The experimental analysis compares the legacy baseline with HMC-IMF across operational and architectural metrics. Interface complexity is measured as a weighted score combining number of direct dependencies, custom transformations, undocumented mappings, and owner fragmentation. The legacy baseline is normalized to 100. HMC-IMF reduces this score to 62 through API reuse, event decoupling, and standardized integration content. Average integration latency falls from 420 ms to 245 ms because frequently used synchronous APIs are routed through governed API management and several non-critical processes are converted to asynchronous event patterns. Mean time to recovery declines from 155 minutes to 54 minutes because alerts include interface ownership, failed message class, runtime layer, and recent deployment history.

TABLE III. EXPERIMENTAL MODERNIZATION OUTCOMES

| Metric | Legacy Baseline | HMC-IMF | Improvement Driver |
|---|---|---|---|
| Interface complexity | 100 | 62 | API reuse and event decoupling |
| Average latency | 420 ms | 245 ms | Pattern-based routing |
| Mean time to recovery | 155 min | 54 min | Telemetry and ownership metadata |
| Failed message rate | 3.8% | 1.1% | Standardized exception handling |
| API onboarding time | 19 days | 6 days | Reusable products and policies |
| Policy coverage | 47% | 91% | Central governance model |
| Interface reuse rate | 22% | 64% | Catalog and API productization |

*Source/derivation note: Derived from the synthetic 24-month operational log evaluation and baseline comparison.*

Failed message rate falls from 3.8 percent to 1.1 percent. The improvement is driven by standardized exception handling, retry policies, monitoring dashboards, and event replay support. API onboarding time declines from 19 days to 6 days because reusable policies, developer products, and documentation templates reduce manual setup. Governance policy coverage increases from 47 percent to 91 percent. Table III reports the full outcome comparison, while Fig. 4 shows the normalized modernization gains. The results should be interpreted as synthetic evaluation outcomes, not production claims. They are intended to demonstrate internal consistency and a plausible magnitude of improvement when integration design, runtime placement, and governance controls are addressed together.

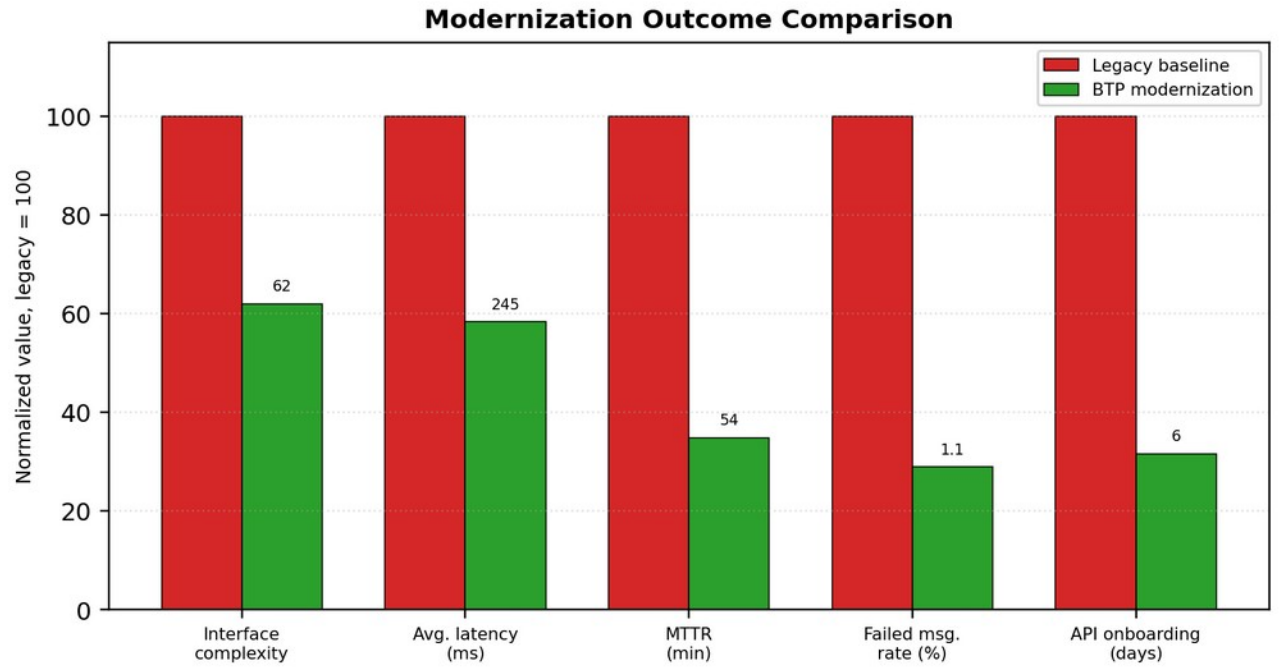

Fig. 4. Modernization outcome comparison, normalized to the legacy baseline where lower values indicate better performance.
Source/derivation note: Derived from Table III performance metrics.

Ablation analysis was also performed. Removing event-driven modernization increases failed message rate from 1.1 percent to 1.7 percent. Removing API governance increases onboarding time from 6 days to 11 days and reduces policy coverage from 91 percent to 74 percent. Removing observability raises mean time to recovery from 54 minutes to 96 minutes. Removing clean-core extension routing increases interface complexity from 62 to 78. These findings suggest that the modernization benefit is distributed across the framework. SAP BTP alone is not sufficient; value depends on disciplined use of integration patterns, security controls, runtime selection, and operational telemetry.

## VI. DISCUSSION

The analysis indicates that SAP BTP can function as a strategic integration modernization layer when implementation decisions are made through an explicit architecture framework. The most important design shift is not simply replacing an old middleware product with a cloud service. It is the movement from interface construction to integration product management. APIs require lifecycle ownership, event topics require governance, extensions require runtime placement discipline, and integration flows require monitoring. The proposed framework brings these concerns into one decision model.

Several practical limitations remain. First, the dataset is synthetic and cannot reflect every enterprise landscape. Real programs face acquisition-specific customizations, regional regulatory differences, undocumented interfaces, inconsistent naming, and incomplete operational telemetry. Second, migration to SAP BTP does not remove the need for strong enterprise architecture governance. Without policy enforcement, a new platform can become another ungoverned integration layer. Third, cost models were not deeply evaluated. Hyperscaler egress, event broker licensing, API traffic growth, and monitoring storage can materially affect the target architecture. Fourth, organizational readiness is essential. Integration modernization requires interface ownership, product teams, security architects, and platform operations working together.

The practical implication is that modernization should be sequenced by value and risk. Interfaces with high failure rates, high business criticality, duplicated mapping logic, or weak security controls should be prioritized before low-volume stable batch flows. Clean-core principles should guide extension decisions, but some legacy interfaces may remain temporarily during phased transition. A realistic roadmap should therefore combine quick wins, governance foundations, platform operating model design, and measured migration waves.

## VII. CONCLUSION AND FUTURE WORKS

This paper presented HMC-IMF, a Hybrid Multi-Cloud Integration Modernization Framework for SAP BTP-centered enterprise architecture. The framework integrates inventory profiling, pattern selection, runtime placement, security policy mapping, observability, and modernization prioritization. A synthetic enterprise dataset with 186 applications, 2,850 interfaces, 1,120 APIs, 740 event topics, 320 batch flows, and 24 months of operational logs was used to evaluate the approach. Compared with a legacy point-to-point baseline, the proposed architecture reduced normalized interface complexity from 100 to 62, average latency from 420 ms to 245 ms, mean time to recovery from 155 minutes to 54 minutes, failed message rate from 3.8 percent to 1.1 percent, and API onboarding time from 19 days to 6 days.

The key conclusion is that SAP BTP modernization should be treated as an enterprise architecture program rather than a tool deployment. SAP Integration Suite, API Management, Event Mesh, Open Connectors, Integration Advisor, Cloud Foundry, Kyma, and clean-core extension practices can produce measurable improvements when deployed through a governed modernization model. Future work should validate the framework using anonymized operational data from real SAP landscapes, extend the prioritization model with cost estimation, incorporate automated security testing for APIs and events, and explore generative AI support for integration documentation and mapping recommendations. Additional research should evaluate how SAP Business Data Cloud and process-mining telemetry can enrich integration observability and support autonomous remediation while maintaining auditability and human control over high-risk changes.

## REFERENCES

[1] SAP, "SAP Business Technology Platform," 2026. [Online]. Available: https://www.sap.com/products/technology-platform.html
[2] SAP, "SAP Integration Suite," 2026. [Online]. Available: https://www.sap.com/products/technology-platform/integration-suite.html
[3] SAP Help Portal, "Migrating from the Neo Environment to the Multi-Cloud Foundation," 2026. [Online]. Available: https://help.sap.com/docs/btp/
[4] F. Liu, J. Tong, J. Mao, R. Bohn, J. Messina, L. Badger, and D. Leaf, "NIST Cloud Computing Reference Architecture," NIST Special Publication 500-292, 2011.
[5] Cloud Native Computing Foundation, "Cloud Native Definition," 2026. [Online]. Available: https://www.cncf.io/about/who-we-are/
[6] A. Wiggins, "The Twelve-Factor App," 2017. [Online]. Available: https://12factor.net/

[7] National Institute of Standards and Technology, "The NIST Cybersecurity Framework (CSF) 2.0," NIST CSWP 29, 2024, doi: 10.6028/NIST.CSWP.29.
[8] ISO/IEC, "ISO/IEC 27001:2022 Information security, cybersecurity and privacy protection, Information security management systems, Requirements," Geneva, Switzerland, 2022.
[9] SAP, "Clean core extensibility for SAP S/4HANA Cloud," 2024. [Online]. Available: https://www.sap.com/documents/2024/09/20aece06-d87e-0010-bca6-c68f7e60039b.html
[10] SAP Help Portal, "Working with API Management," SAP Integration Suite, 2026. [Online]. Available: https://help.sap.com/docs/integration-suite/
[11] SAP, "Open Connectors," SAP Business Technology Platform Use Case, 2026. [Online]. Available: https://www.sap.com/products/technology-platform/use-cases/open-connectors.html
[12] SAP Help Portal, "What Is SAP Event Mesh?" 2026. [Online]. Available: https://help.sap.com/docs/event-mesh/
[13] SAP Help Portal, "Overview of SAP Integration Advisor," SAP Cloud Integration, 2026. [Online]. Available: https://help.sap.com/docs/cloud-integration/
[14] W. Pourmajidi, L. Zhang, J. Steinbacher, T. Erwin, and A. Miranskyy, "A Reference Architecture for Governance of Cloud Native Applications," arXiv:2302.11617, 2023.
[15] A. Berti, G. Park, M. Rafiei, and W. M. P. van der Aalst, "An Event Data Extraction Approach from SAP ERP for Process Mining," arXiv:2110.03467, 2021.
[16] M. Fowler and J. Lewis, "Microservices," 2014. [Online]. Available: https://martinfowler.com/articles/microservices.html
[17] S. Newman, Building Microservices, 2nd ed. Sebastopol, CA, USA: O'Reilly Media, 2021.
[18] S. K. Bhuram and S. Dameruppula, "Enterprise Integration Modernization with SAP BTP: A Strategic Framework for Hybrid and Multi-Cloud Architectures," International Journal of Computational and Experimental Science and Engineering, vol. 12, no. 1, pp. 560-571, 2026, doi: 10.22399/ijcesen.4865.
[19] S. K. Bhuram and S. Dameruppula, "Modernizing Complex Enterprise Landscapes: A Cross-Industry Framework for Legacy System Transformation," Journal of Computational Analysis and Applications, vol. 35, no. 1, pp. 1009-1022, 2026, doi: 10.48047/jocaaa.2026.35.01.81.
[20] S. Dameruppula and S. K. Bhuram, "An AI-Based Framework for Smarter and Faster Data Migration in SAP S/4HANA," Journal of Computational Analysis and Applications, vol. 35, no. 1, pp. 1139-1159, 2026, doi: 10.48047/jocaaa.2026.35.01.93.
[21] S. K. Bhuram, "Process Automation and System Optimization in Enterprise ERP Ecosystems: A Multi-Layer Framework," Computer Fraud and Security, vol. 2026, no. 1, 2026, doi: 10.52710/cfs.921.
[22] S. K. Bhuram, "AI-Driven Secure Data Provisioning for ERP Quality Environments in the Cloud: Architecture, Implementation, and Impact on Automated Financial Reconciliation," Journal of Information Systems Engineering and Management, vol. 11, no. 1S, pp. 1295-1310, 2026, doi: 10.52783/jisem.v11i1s.14278.
[23] G. B. Komarina and S. Dameruppula, "SAP Business Data Cloud Innovation Key Components of SAP Business Data Cloud: Modernize SAP BW, Datasphere, SAP Analytics Cloud (SAC)," The Seybold Report, vol. 20, no. 8, pp. 202-220, Aug. 2025, doi: 10.5281/zenodo.16931168.
[24] S. Dameruppula and S. K. Bhuram, "Digital Transformation of Aviation Supply Chains: An SAP-Based Control Tower Implementation Framework," International Journal of Computational and Experimental Science and Engineering, vol. 12, no. 1, pp. 624-634, 2026, doi: 10.22399/ijcesen.4877.
[25] S. Dameruppula, "The Intersection of Cybersecurity and Privacy in the Digital Age," International Journal of Scientific Research in Computer Science, Engineering and Information Technology, vol. 11, no. 2, pp. 2590-2600, Mar. 2025, doi: 10.32628/CSEIT25112720.